\documentstyle[11pt,newpasp,twoside,epsf]{article}
\def\edcomment#1{\iffalse\marginpar{\raggedright\sl#1\/}\else\relax\fi}
\reversemarginpar

\begin{document}
\title{Geodetic Precession in PSR B1913+16}
\author{Michael Kramer}
\affil{U. of Manchester, Jodrell Bank Observatory, Cheshire,
SK11~9DL, UK}
\author{Oliver L\"ohmer, Aris Karastergiou}
\affil{MPI f\"ur Radioastronomie, 53121 Bonn, Germany}

\begin{abstract}
We review the observational evidence for geodetic precession in PSR
B1913+16 and present the latest observations and results from
modelling the system geometry and beam. 
\end{abstract}

\section{Introduction}

After the discovery of PSR B1913+16 by Hulse \& Taylor in 1974 (Hulse
\& Taylor 1975), it was immediately realized that this system
represents a highly stable and accurate clock orbiting in the
gravitational field of a compact star.  PSR B1913+16 has indeed
fulfilled all promises by finally allowing tests of theories of
gravity in the strong field limit which cannot be realized in the
solar system (see Weisberg, these proceedings). Indeed, no participant
at this conference needs to be reminded about the role which this
pulsar has played in the confirmation of the existence of
gravitational waves. While these tests are based on timing
observations, Damour \& Ruffini (1974) pointed out very soon after the
discovery of this pulsar, that by studying also its emission
properties one can test another prediction made by general relativity,
ie.~that of geodetic precession.

In general relativity, the proper reference frame of a freely falling
object suffers a precession with respect to a distant observer, called
geodetic precession. In a binary pulsar system this geodetic
precession leads to a relativistic spin-orbit coupling, analogous of
spin-orbit coupling in atomic physics.  As a consequence, the pulsar
spin precesses about the total angular momentum, changing the relative
orientation of the pulsar towards Earth. Due to such a change in
geometry, we should also expect a change in the radio emission
received from the pulsar.

The precession rate (e.g.~Boerner et al.~1975) depends on the period
and the eccentricity of the orbit as well as the pulsar and companion
mass. As Joel Weisberg demonstrates in these proceedings, all these
values can be obtained accurately from timing observations. With these
we obtain a precession rate of $\Omega_p=1.21$ deg yr$^{-1}$.  Since
the orbital angular momentum is much larger than the pulsar spin, the
orbital spin practically represents a fixed direction in space,
defined by the orbital plane of the binary system. Given the
calculated precession rate, it takes 297.5 years for the pulsar spin
vector to precess around it.  As a result of the precession the angle
between the pulsar spin axis and our line-of-sight should change with
time, so that different portions of the emission beam are
observed. Consequently, one expects changes in the measured pulse
shape, in particular in the profile width, as a function of time. In
the extreme case, the precession may move the beam out of our
line-of-sight and the pulsar may disappear from the sky until it
becomes visible again.

\section{Previous Studies}

The pulse profiles were naturally studied closely in order to detect
possible changes. Finally, Weisberg et al.~(1989, hereafter
WRT89), discovered a change in the relative amplitude of
the two prominent profile components. While these changes can indeed
be considered as the first signs of the effects of geodetic
precession, a change in the component separation or profile width as
expected from a cone-like pulsar beam was not detected.

Due to precession, the distance of our line-of-sight to the magnetic
axis should also change with time, so that a change in the position
angle (PA) swing of the linearly polarised emission component is
expected.  Cordes, Wasserman \& Blaskiewicz (1990, hereafter
CWB90) studied polarisation data to compare profiles and
PA swings obtained from 1985 to 1988.  CWB90 did neither detect very
clear changes in the pulse shape, nor could they find any significant
change in the PA swing. CWB90 pointed out, however, that the existence
of a core component, which is very prominent at lower
frequencies, complicates the interpretation of the
polarisation data. They noted similar to WRT89 that the core may also
be responsible for the change in relative component
amplitude with time.

PSR B1913+16 is also monitored with the 76-m Lovell telescope and with
the 100-m Effelsberg telescope.  The analysis of Effelsberg profiles
measured between 1994 and 1998 by Kramer (1998) revealed
that the profile components were still changing their relative
amplitude, consistent with the rate first determined by WRT89. Even
more interesting, however, was the first detection of changes in the
separation of the components.  In order to model this long-expected
decreasing width of the profile, two simple assumptions were made,
i.e.~those of a circular hollow cone-like beam and the precession rate
as predicted by general relativity. Both assumptions lead to a model
which has only four free parameters: the misalignment angle $\lambda$
between the pulsar spin and the orbital angular momentum, the
inclination angle between the pulsar spin axis and its magnetic axis,
$\alpha$, the radius of the emission beam, $\rho$, and the precession
phase given by the reference epoch $T_0$.

With the post-Keplerian parameters measured by pulsar timing, general
relativity allows one to compute the value of $\sin i$, i.e.~the sine
of the orbital inclination angle.  For PSR B1913+16, we compute a
value of $i=47.^\circ 2$, or equivalently $i=180-47.2=132.^\circ 8$
whereas the ambiguity cannot be resolved from timing alone.  The best
fit of this model therefore allows four equivalent solutions. One pair
of solutions corresponds to $i=47.^\circ 2$, the other pair to
$i=132.^\circ 8$, respectively.  The remaining choice is given by the
unknown relative orientation of the pulsar spin and the orbital
angular momentum, i.e.~as to whether the pulsar rotation is pro-grade
or retro-grade. It can be argued that a retro-grade case is less
likely (Kramer 1998), so that the polarisation information can be used
to separate the remaining two solutions, as only one gives the correct
observe sense of PA swing.  The finally obtained misalignment angle of
$\lambda={22^{+3}_{-8}}^\circ$ obtained by Kramer (1998) is in
excellent agreement with earlier simulations by
Bailes (1988) who studied the effects of asymmetric
supernova explosions and predicted $\lambda\approx20^\circ$ as a
typical value for PSR B1913+16-like systems.

The obtained best fit also lead to the prediction that the pulsar will
disappear from the sky around the year 2025! Moreover, it also implies
that the component separation remains almost unchanged for about 60
yr. It is now easy to understand why WRT89 were not lucky to detect
changes in the component separation. Similarly, computing the change
in PA swing which had to be measured by CWB90 for a positive detection
of a geometry change, produces a value which is only slightly larger
than their estimated detection limit. It should also be noted that
based on an emission model and the relative change in component
ratio alone, Istomin (1991) also suggested a disappearance
of the pulsar around 2020. The full model as presented
by Kramer (1998) also predicts a reappearance around the year
2220. PSR B1913+16 will, in total, only be observable for about a
third of the precession period. While this seems to affect possible
detection rates of double neutron star systems and hence the detection
rate of gravitational wave detectors like LIGO or GEO600, averaged
over time existing numbers do not change, as discussed
in more detail by Kramer (1998, 2002).

\begin{figure}[hbt]
\plotone{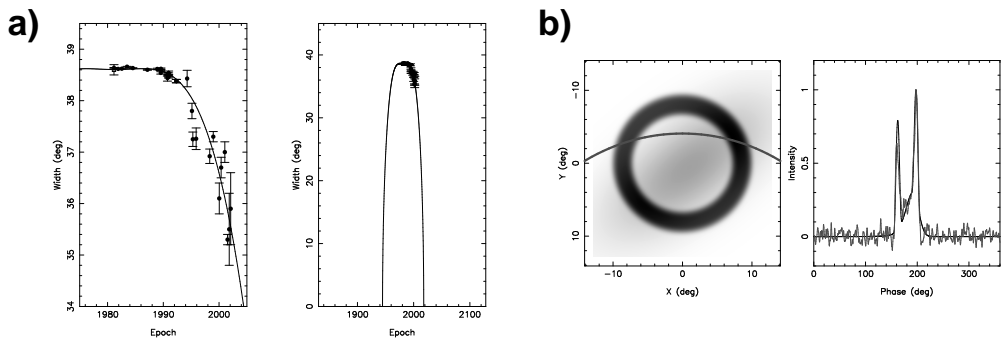}
\caption{
(a): Measurements of component separation as
a function of time showing epochs from 1982 to 2002 (left)
and for full precession cycle (right). Figure includes
some early data taken by WRT89 and WT02. (b): Beam model
including an off-set core component (left) and comparison
of model profile with data (right).}
\end{figure}

\section{Recent Results and Update}

In Figure 1a we show the latest measurements for the component
separation as a function of time, demonstrating that the profile
continues to narrow as predicted by the model.  Weisberg \& Taylor
(2000, 2002, also these proceedings) also obtained new
measurements after the Arecibo upgrade, confirming the results
reported above. Thanks to the superior sensitivity of the
telescope, they did not only measure the same decrease in component
separation, but could also measure a general decrease in profile width
at several intensity levels. Using these data they derive a
geometry which is in agreement with that of Kramer (1998), and
they also obtain a map of a pulsar emission beam for the first time.
Since our line-of-sight moves through the emission beam, each profile
represents a slightly different cut through the beam structure.
During their data analysis, Weisberg \& Taylor separate the measured
profiles into odd and even parts and use the width information for all
intensity levels of the even profiles, combined with a mapping
function, to derive a model for geometry and beam shape.  The results
of this mapping process are surprising as the beam seems not only to be
elongated but even hour-glass shaped.

We use our data for an complementary, alternative approach. We propose
to use the original profiles in order to maintain information about
the features causing the profile asymmetry, namely the off-set core.
We then compute profiles observable at different epochs taking full
spherical geometry into account, and compare these model profiles to
the observed data (see Figure 1b). While this work is still in
progress, initial results suggest that the beam may indeed be slightly
elongated although an hour-glass beam shape may not be necessary to
explain the data.  Further modelling will be necessary but it may
provide us with a beam map which can then be used as input for tests
of the precession rate. The observations of other precessing pulsars
like PSR B1828$-$11 (Stairs et al.~2000) or perhaps PSR B1931+24
(Kramer et al.~in prep.) may help in this process to understand the
pulsar beam pattern. At the moment, the detection of effects of
geodetic precession in PSR B1913+16 is a successful qualitative test
of general relativity, but it may become possible to perform also a
quantitative test by measuring the precession rate using our beam
models.

\acknowledgements
We thank Joel Weisberg and Norbert Wex for many useful discussions. 
We are also grateful to everyone helping with the Effelsberg observations
over the years.

\end{document}